\documentclass[letterpaper]{rmaa}

\usepackage{longtable}

\usepackage{paralist}

\usepackage{psfrag,color}

\usepackage[latin1]{inputenc}

\newcommand{\eb}{\begin{equation}}
\newcommand{\ee}{\end{equation}}
  
\title{Absolute N\lowercase{uv} magnitudes of Gaia DR1 astrometric stars and a search for hot companions in nearby systems} 

\author{
  Valeri V. Makarov\affil{US Naval Observatory, 3450 Massachusetts Ave NW, Washington DC 20392-5420, USA}}

\shortauthor{Makarov}
\shorttitle{Absolute Nuv magnitudes of nearby stars}

\fulladdresses{
\item Valeri V. Makarov: US Naval Observatory, 3450 Massachusetts Ave NW, Washington DC 20392-5420, USA (valeri.makarov@navy.mil).}

\SetYear{2017}
\SetVolume{53}
\ReceivedDate{May 2 2017}
\AcceptedDate{June 23 2017}

\listofauthors{V. V. Makarov}
\indexauthor{Makarov, V. V.}

\abstract{Accurate parallaxes from Gaia DR1 (TGAS) are combined with GALEX visual Nuv magnitudes to produce absolute Mnuv magnitudes and
an ultraviolet HR diagram for a large sample of astrometric stars. A functional fit is derived of the lower envelope main sequence of
the nearest 1403 stars (distance $< 40$ pc), which should be reddening-free. Using this empirical fit, 50 nearby stars are selected with
significant Nuv excess. These are predominantly late K and early M dwarfs, often associated with X-ray sources, and showing other manifestations of magnetic activity. The sample may include systems with hidden white dwarfs, stars younger than the Pleiades, or, most likely, 
tight interacting binaries of the BY Dra-type. A separate collection of 40 stars with precise trigonometric parallaxes and
Nuv$-$G colors bluer than 2 mag is presented. It includes several known novae, white dwarfs, and binaries with hot subdwarf (sdOB) components, but
most remain unexplored.}

\resumen{
Se combinan paralajes del Gaia DR1 (TGAS) con magnitudes visuales Nuv del GALEX para obtener magnitudes absolutas 
Mnuv as\'i como un diagrama HR ultravioleta para una muestra de estrellas astrom\'etricas. Se deriva un ajuste  
para la envolvente inferior de la secuencia principal para las 1403 estrellas con distancias $< 40$~pc, que no 
tendr\'ian enrojecimiento. Se seleccionan 50 estrellas cercanas con un exceso Nuv 
considerable. Estas estrellas son principalmente de tipo K tard\'io y M temprano, frecuentemente asociadas con fuentes de rayos 
X, y con otras manifestaciones de actividad magn\'etica. La muestra puede incluir sistemas con enanas blancas ocultas, 
es\-tre\-llas m\'as j\'ovenes que las Pl\'eyades o binarias interactuantes de tipo BY Dra. Se 
presenta otra muestra de 40 es\-tre\-llas con paralajes trigonom\'etricas precisas y colores Nuv-G m\'as azules que 2~mag. \'Esta 
incluye varias novas, enanas blancas y binarias con subenanas calientes como compa\~neras.
}

\addkeyword{binaries: general}
\addkeyword{Hertzsprung--Russell and C--M diagrams}
\addkeyword{stars: activity}
\addkeyword{white dwarfs}

\begin{document}
\maketitle

\section{Introduction}
\label{intro.sec}

The first release of the Gaia mission data (Gaia DR1) includes two astrometric catalogs \citep{bro}. The smaller catalog, called
TGAS, includes 2 million brighter stars with accurate proper motions and parallaxes and is based on a combination of astrometric
data from Hipparcos and Tycho-2 \citep{esa,hog} and Gaia itself \citep{lin}, while the larger catalog of 1.1 billion objects
is derived from Gaia's own observations and ICRF-2 radio source positions. I am using the TGAS in this paper, specifically,
the parallaxes of brighter stars listed there. The formal errors of parallaxes are all smaller than 1 mas, which was the only
requirement for an astrometric solution to be included in DR1. The entire set of 2 million Gaia DR1 stars was cross-matched
with the GALEX DR5 catalogs by \citet{bia}, namely, the All-Sky Imaging survey (AIS) with  limiting magnitudes 19.9/20.8 in FUV/NUV
and the Medium-depth Imaging Survey (MIS) with limiting magnitudes 22.6/22.7. The search for Galex matches was performed with Gaia
J2015 positions in a cone of $3.5\sigma$ of Galex positions, but not greater than $5\arcsec$ on the sky. The total number of matched
sources is $720\,622$, which is a surprisingly high rate given that the GALEX catalog covers only a little more than half of the sky\footnote{
There is a wide empty swath in GALEX along the Galactic plane due to the source confusion.}. GALEX DR5 provides precise far-ultraviolet
(Fuv; 1344-1786 $\AA$) and near-ultraviolet (Nuv; 1771-2831 $\AA$) magnitudes with errors generally about 0.02--0.03 mag. We thus obtain
a large collection of astrometric standards with good parallaxes and UV magnitudes which can be used to compute {\it absolute}
ultraviolet magnitudes:
\eb
M{\rm nuv}={\rm Nuv}-10+5\log \varpi,
\ee
where $\varpi$ is the parallax in mas. The uncertainty of absolute magnitudes is dominated by the error of Nuv magnitude for most of the
stars, but distant objects (small parallaxes) can have the ratio $\varpi/\sigma_\varpi$ close to unity, to the point that the observed
parallax takes a negative value. To reduce the astrometric noise component in the subsequent analysis, the sample needs to be
limited to the most reliable determinations with large $\varpi/\sigma_\varpi$ or, which is almost equivalent in this case, with large
parallaxes.

\vspace{0.3cm}

\section{The Hertzsprung--Russell diagram in the near-ultraviolet}
\label{dia.sec}

\begin{figure}[!t]\centering
\includegraphics[width=\columnwidth]{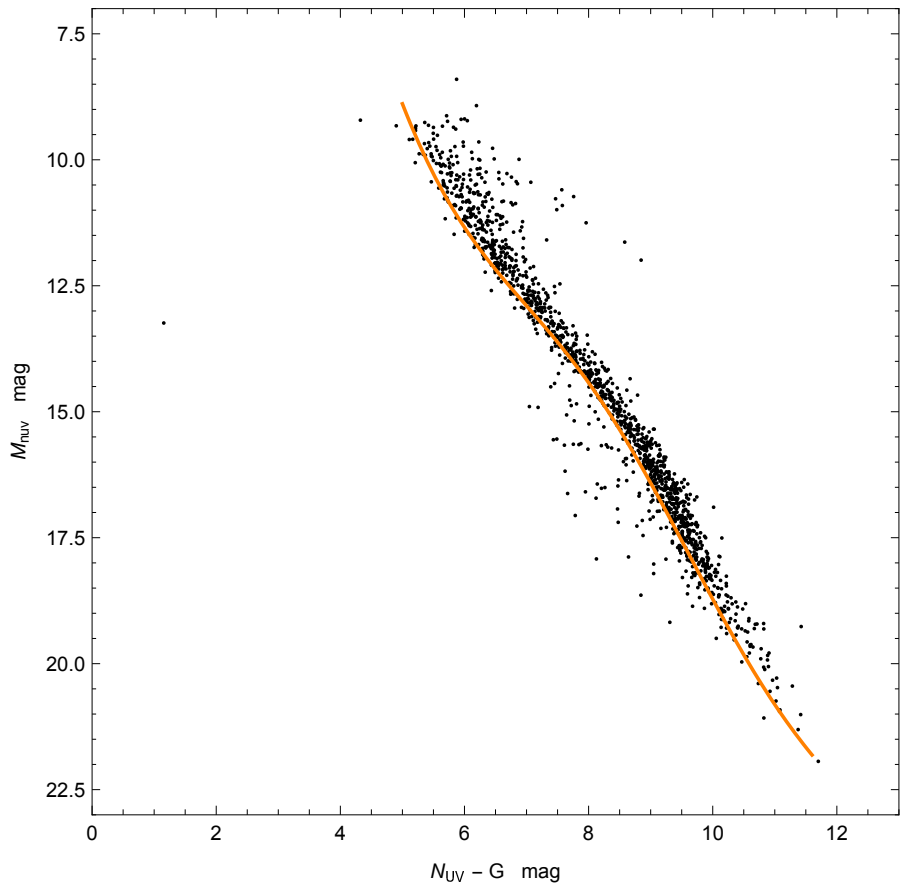}
\caption{Near-UV HR diagram of absolute Nuv magnitudes versus Nuv$-G$ colors for nearby stars with parallaxes greater than 25 mas,
matched with GALEX sources. The single dot far to the left represents the only single white dwarf in the sample, DN Dra.
The line along the lower envelope of the main sequence shows the formal functional fit, see text.}
  \label{dia.fig}
  \vspace{0.9cm}
\end{figure}

Figure \ref{dia.fig} displays the ``absolute Nuv magnitude versus Nuv$-G$ color" (HR) diagram for 1403 stars selected with the logical {\it and},
or intersection, of the following criteria: $\varpi/\sigma_\varpi>5$; $\varpi>25$ mas. Although observed $B_T$ and $V_T$ magnitudes
are available for all Tycho-2 stars, as well as derived Johnson $B$ and $V$ magnitudes, I will use the more accurate broadband $G$
magnitudes as observed by Gaia. For the sample under consideration, the distribution of formal errors of $G$ magnitudes peaks
at 0.0005 mag with a median of 0.0009 mag. This is much smaller than the uncertainty of Nuv magnitudes.
The selection includes stars confidently within 40 pc of the Sun.

Most of the stars lie on a well-defined and narrow main sequence stretching between magnitudes 9 and 22 in absolute magnitude $M$nuv
and 5 -- 12 in Nuv$-G$ color. There is a rudimentary giant branch veering off to the right at the top of the main sequence, reflecting
the scarcity of giants in the immediate solar neighborhood. The width of the main sequence is roughly 0.5 mag, likely to come from
unresolved binaries. A regular MS--MS binary is shifted up and to the right of the main sequence because there is more
additional flux in the $G$ band than in Nuv. The largest deviation from the main sequence due to binarity is $\approx 0.75$ mag
for identical twin pairs. Interstellar reddening is not expected to have a significant presence in this diagram as there are no dense
dust-molecular clouds within 40 pc. 

Using a standard nonlinear fit algorithm (minimizing the residual RMS), 
this functional form is found for the \emph{lower envelope} main sequence, represented with
a solid curved line in Figure \ref{dia.fig}:
\begin{eqnarray}
M{\rm nuv}({\rm fit})& = & 15.339 + 5.708\,x - 1.653\,x^2 + \nonumber \\
& & 1.029\,x^3 - 0.915\,\cos[\pi x],
\label{fit.eq}
\end{eqnarray}
where $x=({\rm Nuv}-G-8)/3.3$. This curve accurately represents the bluest magnitudes and colors of normal field dwarfs without any signs
of activity or reddening. The presence of a cosine term is justified by the much better results achieved: the standard deviation of
post-fit residuals over a set of nodal points goes down from 0.21 to 0.08 mag, and the observed wiggles of the lower envelope
are much more truthfully represented. 
A median main sequence functional expansion can be obtained from Equation \ref{fit.eq} by replacing the constant
term 15.339 with 15.150.

\vspace{0.3cm}

The inverse main sequence fit, i.e., $[{\rm Nuv}-G](M{\rm nuv})$ may be handy if we want to estimate the amount of observed ``UV-excess" for a
known absolute magnitude:
\begin{eqnarray}
[{\rm Nuv}-G]({\rm fit}) & = & 8.096 + 3.463\,y + 0.786\,y^2 - \nonumber \\
& & 0.120\,y^3 + 0.482\,\cos[\pi y],
\end{eqnarray}
where $y=(M{\rm nuv}-15.5)/6.5$.

\vspace{0.2cm}

\begin{figure}[!t]\centering
\includegraphics[width=\columnwidth]{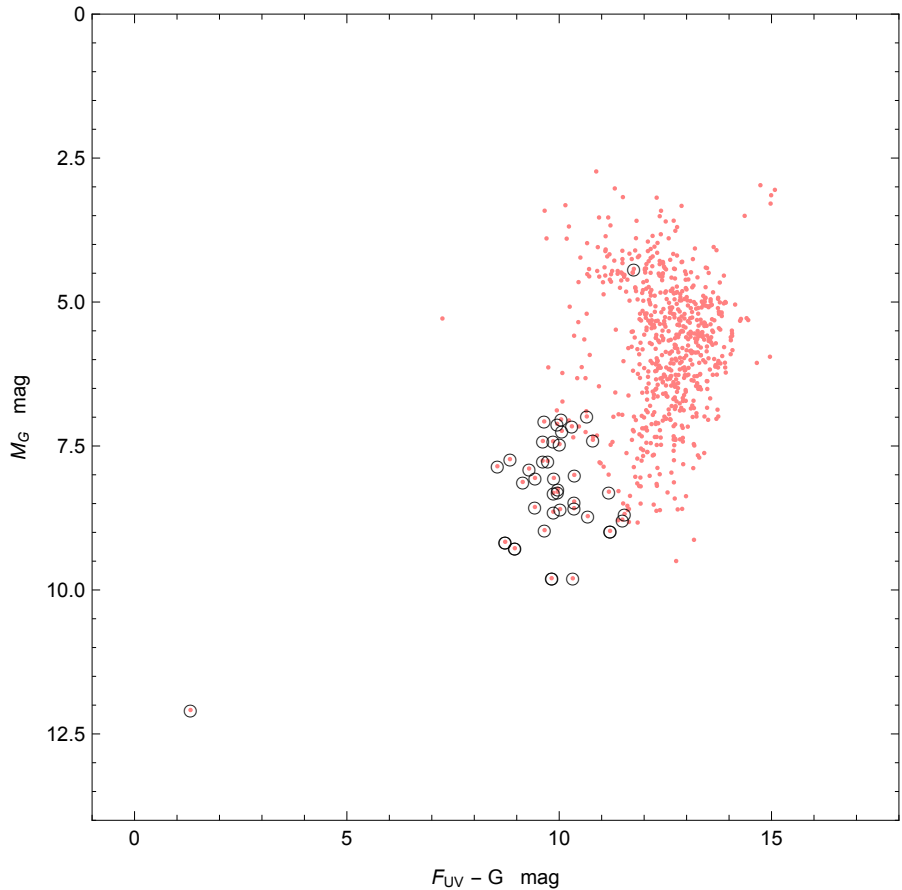}
\caption{Far-UV HR diagram of absolute $G$ magnitudes versus Fuv$-G$ colors for nearby stars with parallaxes greater than 25 mas,
matched with GALEX sources. The single encircled dot in the lower left corner represents the only single white dwarf in the sample, DN Dra.
Stars of significant Nuv luminosity excess collected in Table 1 are also marked with circles.}
  \label{fuv.fig}
  \vspace{0.3cm}
\end{figure}

A few dozen stars lie to the left of the lower boundary curve with either their colors too blue or absolute magnitudes too faint. The latter
is unlikely because of the slope of the main sequence -- a deficit of Nuv flux would shift the point to the right of the main sequence.
Hence, the stars below and to the left of the main sequence envelope have ultraviolet excess with respect to normal luminosities.
This is confirmed by Figure~\ref{fuv.fig} which shows a similar HR diagram for $M_G$ for the same set of stars, versus \mbox{Fuv$-+G$} color. The 
main sequence is not well defined in this diagram, but the stars with a large Nuv luminosity excess occupy a specific area of the diagram
with absolute $M_G$ magnitudes greater than 6.7 and Fuv$-G$ colors less than 11.6. This confirms that the ultraviolet excess for a fraction 
of nearest dwarfs is real and present in a wide range of wavelengths.

\vspace{0.3cm}

\section{Nearby stars with Nuv excess}
\label{nuv.sec}
Table 1 lists nearby stars (distance less than 40~pc) found in the TGAS--GALEX sample with significant Nuv luminosity excess. The latter
was defined as $M{\rm Nuv}$(obs)$-M{\rm Nuv}$(fit)$>0.7$ mag. This is a conservative limit possibly leaving out a number of genuine
sources of enhanced UV radiation, but it results in a more manageable sample of 50 stars which can be
individually verified. The columns of the table include: (1) RA J2015 in degrees; (2) Dec J2015 in degrees; (3) HIP number when available;
(4) Tycho-2 identification when HIP number is not available; (5) parallax in mas; (6) standard error of parallax in mas; (7) $G$ magnitude;
(8) Fuv magnitude, if available; (9) Nuv magnitude; (10) formal error of Fuv magnitude, if available; (11) formal error of Nuv magnitude.
Columns 1 through 7 are copied from TGAS, while Columns 8 -- 11 are copied from GALEX.

\begin{table*}[!t]\centering
\setlength{\tabnotewidth}{\columnwidth}\tablecols{11}
  \setlength{\tabcolsep}{0.89\tabcolsep}
  \caption{Nearby Dwarfs (distance $<$ 40 pc) with excess N\lowercase{uv} luminosity}\label{1.tab}
\begin{tabular}{rrccccccccc}
    \toprule
\multicolumn{1}{c}{(1)} & \multicolumn{1}{c}{(2)} & (3) & (4) & (5) & (6) & (7) & (8) & (9) & (10) & (11) \\
\multicolumn{1}{c}{RA J2015} & \multicolumn{1}{c}{Dec J2015} & HIP & Tycho-2 & $\varpi$ & $\sigma_\varpi$ & $G$ & Fuv & Nuv & Fuv sig & Nuv sig \\
\multicolumn{1}{c}{$\deg$} & \multicolumn{1}{c}{$\deg$} & & & mas & mas& mag & mag & mag& mag & mag\\ 

    \midrule
 51.028 & 23.7845 & 15844 &     & 48.59 & 0.31 & 9.809 & 19.779 & 18.072 & 0.206 & 0.053 \\
 35.8615 & 22.7347 & 11152 &     & 36.86 & 0.34 & 10.291 & 19.432 & 17.723 & 0.152 & 0.041 \\
 37.4013 & 34.3948 &     & 2331-1138-1 & 26.29 & 0.91 & 11.46 & 20.888 & 19.081 & 0.298 & 0.079 \\
 10.7024 & 35.5491 & 3362 &     & 46.4 & 0.31 & 9.422 & 19.032 & 17.411 & 0.141 & 0.027 \\
 15.9179 & 40.8574 & 4967 &     & 30.33 & 0.52 & 10.008 & 19.864 & 17.654 & 0.192 & 0.043 \\
 168.974 & 55.3304 &     & 3828-36-1 & 35.03 & 0.29 & 10.332 & 20.206 & 17.82 & 0.239 & 0.029 \\
 169.016 & 52.7767 & 55043 &     & 25.69 & 0.25 & 7.842 &     & 12.165 &     & 0.003 \\
 158.462 & 49.1867 & 51700 &     & 26.44 & 0.28 & 7.313 & 19.071 & 12.214 & 0.111 & 0.003 \\
 97.7541 & 50.046 &     & 3384-35-1 & 49.58 & 0.24 & 10.119 & 20.14 & 18.239 & 0.163 & 0.032 \\
 108.118 & 45.4218 &     & 3392-2038-1 & 25.19 & 0.33 & 10.802 &     & 18.646 &     & 0.063 \\
 139.843 & 62.0531 & 45731 &     & 25.75 & 0.38 & 10.359 & 19.974 & 18.127 & 0.112 & 0.02 \\
 149.623 & 67.054 & 48899 &     & 33.47 & 0.29 & 9.769 & 20.563 & 18.078 & 0.218 & 0.023 \\
 159.943 & 65.7559 &     & 4150-1189-1 & 30.77 & 0.38 & 10.561 & 20.916 & 19.033 & 0.315 & 0.076 \\
 94.5296 & 75.1012 &     & 4525-194-1 & 32.23 & 0.42 & 10.35 & 19.641 & 18.104 & 0.115 & 0.033 \\
 230.471 & 20.9783 & 75187 &     & 86.81 & 0.38 & 8.948 & 18.809 & 16.897 & 0.084 & 0.019 \\
 233.155 & 46.8846 &     & 3483-856-1 & 38.11 & 0.7 & 10.559 & 20.912 & 19.025 & 0.205 & 0.066 \\
 252.108 & 59.0551 & 82257 &     & 91.04 & 0.5 & 12.288 & 13.606 & 13.443 & 0.007 & 0.004 \\
 332.877 & 18.4269 & 109555 &     & 85.75 & 0.3 & 9.112 & 20.599 & 18.624 & 0.228 & 0.06 \\
 274.354 & 48.3675 &     & 3529-1437-1 & 50.28 & 0.88 & 10.211 & 20.884 & 18.688 & 0.278 & 0.062 \\
 279.857 & 69.0518 &     & 4430-329-1 & 30.79 & 0.47 & 10.871 & 20.733 & 19.078 & 0.231 & 0.068 \\
 5.03554 & -17.0614 & 1608 &     & 43.31 & 0.57 & 11.687 &     & 20.996 &     & 0.216 \\
 347.082 & -15.41 & 114252 &     & 39.85 & 0.25 & 10.052 & 19.482 & 17.662 & 0.109 & 0.03 \\
 347.168 & -16.3833 &     & 6395-1046-1 & 26.15 & 0.61 & 9.899 & 20.549 & 17.351 & 0.376 & 0.053 \\
 339.692 & -20.6215 & 111802 &     & 112.68 & 0.38 & 8.035 & 17.996 & 16.175 & 0.051 & 0.014 \\
 353.129 & -12.2646 &     & 5832-666-1 & 36.02 & 0.53 & 9.684 & 19.692 & 17.89 & 0.119 & 0.034 \\
 2.77028 & -5.78394 & 897 &     & 41.88 & 0.91 & 11.129 &     & 19.772 &     & 0.06 \\
 340.293 & -16.4196 &     & 6386-326-1 & 25.01 & 0.98 & 11.501 &     & 20.279 &     & 0.168 \\
 73.1023 & -16.8236 &     & 5899-26-1 & 63.4 & 0.37 & 10.264 & 19.219 & 18.049 & 0.109 & 0.039 \\
 72.491 & -14.2861 &     & 5328-261-1 & 27.87 & 0.32 & 10.527 &     & 19.141 &     & 0.072 \\
 110.931 & 20.4153 &     & 1355-214-1 & 36.12 & 0.31 & 9.369 & 19.668 & 17.089 & 0.139 & 0.028 \\
 203.679 & -8.34242 & 66252 &     & 48.39 & 0.35 & 8.614 & 18.659 & 16.42 & 0.114 & 0.023 \\
 165.66 & 21.9669 & 53985 &     & 83.77 & 0.35 & 8.678 & 19.848 & 17.543 & 0.204 & 0.038 \\
 230.356 & 4.24718 &     & 344-504-1 & 36.44 & 0.84 & 10.872 & 22.411 & 20.118 & 0.453 & 0.107 \\
 259.975 & 26.5023 & 84794 &     & 93.18 & 0.49 & 9.951 & 20.271 & 18.794 & 0.197 & 0.063 \\
 38.5944 & -43.7976 & 11964 &     & 86.14 & 0.32 & 8.052 & 16.892 & 15.238 & 0.03 & 0.009 \\
 0.614199 & -46.0289 & 191 &     & 27.17 & 0.39 & 11.409 & 21.76 & 20.286 & 0.393 & 0.09 \\
 47.0292 & -24.7591 & 14568 &     & 30.74 & 0.4 & 9.635 & 19.282 & 17.331 & 0.125 & 0.035 \\
 28.2985 & -21.0951 &     & 5858-1893-1 & 31.85 & 0.6 & 10.336 & 18.88 & 17.382 & 0.086 & 0.03 \\
 25.8094 & -21.6157 & 8038 &     & 32.35 & 0.85 & 9.566 & 19.518 & 16.956 & 0.101 & 0.016 \\
 45.6603 & -18.1656 & 14165 &     & 52.47 & 0.31 & 10.586 &     & 20.261 &     & 0.107 \\
    \bottomrule
\end{tabular}
\end{table*}
\begin{table*}[!t]\centering
\setlength{\tabnotewidth}{\columnwidth}\tablecols{11}
  \setlength{\tabcolsep}{0.89\tabcolsep}
 TABLE 1 (CONTINUED)
\begin{tabular}{rrccccccccc}
    \toprule
\multicolumn{1}{c}{(1)} & \multicolumn{1}{c}{(2)} & (3) & (4) & (5) & (6) & (7) & (8) & (9) & (10) & (11) \\
\multicolumn{1}{c}{RA J2015} & \multicolumn{1}{c}{Dec J2015} & HIP & Tycho-2 & $\varpi$ & $\sigma_\varpi$ & $G$ & Fuv & Nuv & Fuv sig & Nuv sig \\
\multicolumn{1}{c}{$\deg$} & \multicolumn{1}{c}{$\deg$} & & & mas & mas& mag & mag & mag& mag & mag\\ 

   \midrule
117.301 & -76.7027 &     & 9381-1809-1 & 92.06 & 0.49 & 9.975 & 19.797 & 18.101 & 0.15 & 0.035 \\
130.385 & -68.4272 & 42650 &     & 32.84 & 0.52 & 10.289 &     & 18.771 &     & 0.081 \\
159.939 & -44.5109 &     & 7722-1583-1 & 53.57 & 0.97 & 10.519 & 19.247 & 19.567 & 0.148 & 0.121 \\
 161.298 & -26.1259 &     & 6638-293-1 & 30.3 & 0.6 & 9.965 &     & 18.249 &     & 0.038 \\
 139.085 & -18.6252 &     & 6032-282-1 & 73.35 & 0.94 & 9.647 & 20.844 & 18.696 & 0.319 & 0.097 \\
 201.451 & -28.3744 & 65520 &     & 65. & 0.31 & 9.957 &     & 19.55 &     & 0.114 \\
 349.888 & -39.6569 &     & 8006-520-1 & 25.59 & 0.77 & 11.103 &     & 20.095 &     & 0.105 \\
 341.242 & -33.251 & 112312 &     & 48.17 & 0.57 & 10.547 & 20.203 & 18.209 & 0.142 & 0.036 \\
 311.291 & -31.3424 & 102409 &     & 102.12 & 0.39 & 7.712 & 17.442 & 15.588 & 0.042 & 0.01 \\
 318.272 & -17.4875 &     & 6351-286-1 & 26.91 & 0.53 & 10.087 & 20.148 & 18.092 & 0.171 & 0.035 \\
    \bottomrule
\end{tabular}
\vspace{0.3cm}
\end{table*}    

The single point far to the left in Figures \ref{dia.fig} and \ref{fuv.fig} represents the well-known white dwarf {\bf DN Dra = GJ 1206}
of spectral type DA4.0 \citep[e.g.,][]{fon}. It is very luminous in the near-UV with an absolute magnitude $M$nuv$=7.95$ mag. The absence
of other bright white dwarfs within 40 pc of the Sun in our selection is probably explained by selection effects in the Hipparcos,
Tycho-2, and TGAS catalogs\footnote{\citet{gon} find only 15 WD from Hipparcos and Tycho-2 using their astrometric and photometric
criteria, but only 4 of them are present in TGAS with listed parallaxes and proper motions, namely, HIP 82257 = DN Dra, TYC 3953-480-1 =
Eggr 378, TYC 8942-2593-1, and TYC 1538-1368-1, the latter two being false positives (not WD) because of gross errors in the
Tycho-2 proper motions. The cause of this low representation rate is unknown.}. Other excess stars have much redder Nuv$-G$ colors and cannot 
be isolated white dwarfs. An extensive
literature and astronomical database search with VizieR and Simbad reveals that the sample includes predominantly dwarfs of late K to
early M spectral types. Some of these stars are included in the study of the near-UV luminosity function of early M-type dwarfs
by \citet{ans}, where the authors used Nuv fluxes relative to visual and near-infrared fluxes rather than absolute luminosities, which
leads to a larger sample. \citet{ans} find that up to 1/6 of all such M dwarfs show elevated levels of near-UV radiation, which may be
inconsistent with a constant star-formation rate and commonly used age-activity relations. Here we find a much lower rate of
dwarfs with excess Nuv luminosities in absolute units ($\sim3.6$\%). It is possible that a relative-flux selection is biased toward
more active M dwarfs from a larger volume of space. 

\vspace{0.3cm}

\subsection{Too Many Young Stars?}

All of our late-type dwarfs satisfy the rather generous selection criteria for young stars of \citet[][their Fig. 1]{rod}. Can they all be young?
Assuming a constant rate of star formation over the 13 Gyr history of the Galaxy, the rate of overluminous dwarfs corresponds to a
threshold age of 460~Myr. Hence, the existence of such dwarfs in the solar neighborhood can be explained if stars younger than the Hyades
can retain the observed Nuv excess due to a high level of magnetic activity fueled by fast rotation. There are no star forming regions,
OB associations, or young open clusters within the close solar neighborhood. However, some of the stars listed in Table~1 have been
proposed as members of sparse young moving groups (YMG). Some interesting examples are:
\begin{itemize}
\item
{\bf TYC 5899-26-1}, an M3.3 dwarf, was assigned by \citet{shk12} to the AB Doradus YMG with an estimated age of 30--50 Myr \citep{mak07}.
\item
{\bf TYC 5832-666-1}, a rotationally variable M0 dwarf, was assigned by \citet{lep} to the $\beta$ Pic YMG with an estimated age of 
\mbox{20--30 Myr}.
\item
{\bf HIP 84794 = GJ 669A}, a flaring M3.5 dwarf, was assigned by \citet{shk12} to the Hyades MG with an estimated age of 
600~Myr\footnote{The existence of the Hyades moving group as a coeval aggregate of stars has been disputed, 
\citep[e.g.,][]{fam}.}. 
\item {\bf HIP 112312 = WW PsA}, an M1 dwarf, was assigned by \citet{shk12} to the $\beta$ Pic MG, but it is also a rotationally variable
binary of the BY Dra type.
\item {\bf HIP 102409 = GJ 803 = AU Mic}, a famous M1e young dwarf with a resolved debris disk, considered to be a member of the $\beta$ Pic MG,
but also an active binary of the BY Dra type.
\item
{\bf TYC 6351-286-1 = HD 201919}, a rotationally variable K6Ve dwarf suggested by \citet{ell} as a member of the AB Doradus YMG.
\end{itemize}

\begin{figure}[!t]\centering
\includegraphics[width=\columnwidth]{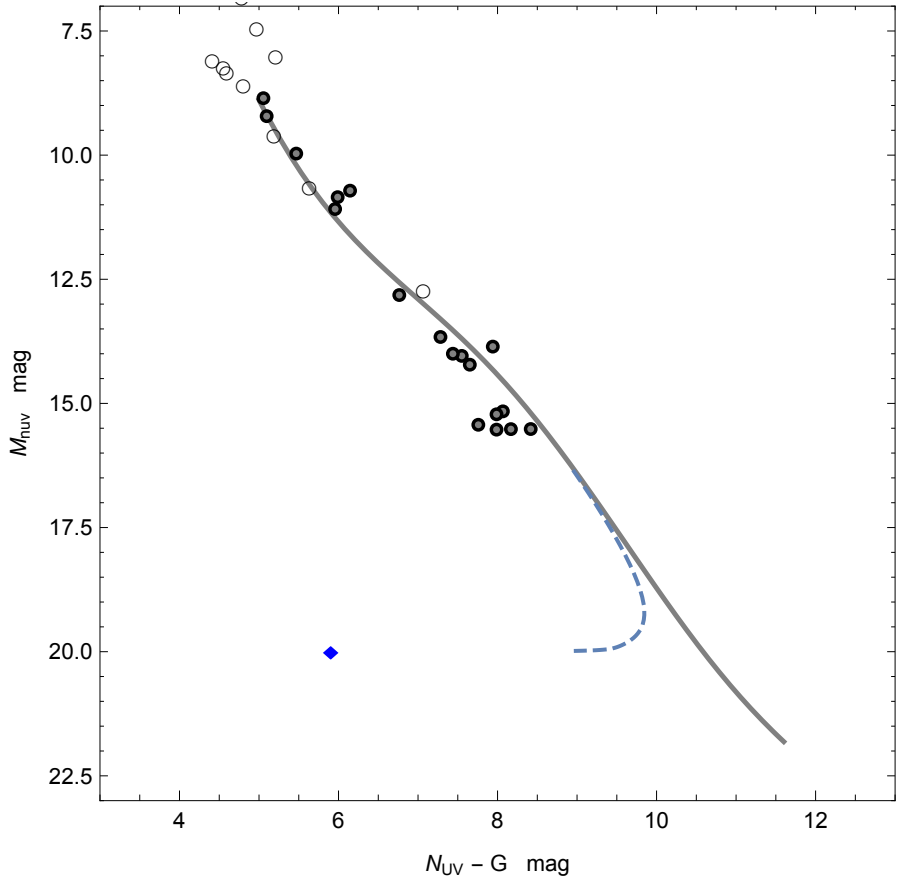}
\caption{Near-UV HR diagram of absolute Nuv magnitudes versus Nuv$-G$ colors for known members of the Pleiades cluster (open circles)
and Tuc Hor MG members (filled disks) found in the TGAS--GALEX cross-match. The diamond in the lower left corner shows the data for
the nearby white dwarf van Maanen 2. The solid line is the main sequence fit for nearby stars (Equation \ref{fit.eq}).
The dashed line shows the computed loci of old inactive dwarfs in unresolved binaries with white dwarfs identical to van Maanen 2.}
  \label{young.fig}
  \vspace{0.3cm}
\end{figure}

\citet{shk14} conclude that the median UV flux of early M stars remains at ``saturated" levels for a few hundred Myr, and the decline
in activity after $\approx 300$ Myr follows a time$^{-1}$ dependence, but their analysis is based on rather rough distance estimates
and the relative $F_{\rm UV}/F_J$ flux ratio. Often, the proposed membership of stars to the nearest moving groups is uncertain
and suffers from considerable rates of interlopers. The earlier attempts at identifying such groups were based on proper motion and X-ray
count rate data following the successful completion of the Rosat and Tycho-2 missions \citep{mak00}. But the census of nearby most
luminous stars in X-rays shows that this criterion nets more active binaries of the RS CVn and BY Dra type than very young objects \citep{mak03}.
Even though the majority of objects in Table~1 are associated with Rosat-detected X-ray sources, this does not guarantee their young age.
Figure \ref{young.fig} presents an attempt to verify that the over-luminous dwarfs can be younger than the Hyades. Only 9 known Pleiades
members seem to be present in the TGAS--GALEX sample, marked with open circles. These stars are solar-type or earlier, and they
conform to the main sequence fit quite well. Unfortunately, small-mass dwarfs are missing, perhaps because they are too faint.
The filled circles represent the proposed members of the nearer and possibly younger Tuc Hor MG (estimated age 27 Myr) from \citet{mak07, kra}.
They allow us to probe later spectral types down to the early K. These candidate young stars start to deviate from the main sequence
at $M$nuv$\approx$14--15 mag. This may be interpreted as a ``turn-on" point of very young stars, which is likely age-dependent.
The absence of late-type dwarfs thwarts verification of this result. The preliminary conclusion is that stars younger than the Pleiades
($\la100$ Myr) that wandered by chance into the close solar neighborhood may be significantly over-luminous in the UV compared to older
inactive field stars, but their number should be much smaller than what we find on the HR diagram.

\vspace{0.1cm}

\subsection{Hidden White Dwarfs}
The selection criteria adopted in \S~\ref{nuv.sec} are sensitive to unresolved binaries that include a cool main sequence dwarf and
a hotter white dwarf (WD). \citet{fuh} speculated that binaries with WD companions should be quite common in the solar neighborhood
but it is not easy to find them on account of their optical dimness. In principle, the near-UV HR diagram method should be
capable of detecting hidden WD companions from the youngest and hottest (but rare) to objects as late as D8, or approximately 6300 K
in effective temperature, but the prospects strongly depend on the spectral type of the main-sequence primary. The easiest and the
most common target would be M dwarfs, and indeed, the prevalence of such objects in Table~1 can be explained this way. The dashed curved
line in Fig.~\ref{young.fig} shows the loci of M--WD pairs with completely blended photometry, where the cool nearby WD 
{\bf van Maanen 2 = GJ 35 = HIP 3829} is used as a WD template. GJ 35 is a very close Population II white dwarf which is missing in TGAS
(but present in the main Gaia catalog) of DZ7.5 spectral type, marked with a diamond on the diagram. Blended MS--WD pairs cannot
be bluer than the WD component or significantly redder than the MS component; thus, their positions are limited to the sharp
angle formed by the main sequence and the horizontal line through the $M$nuv of the WD. No WD companions have been identified
in the literature for stars listed in Table~1 but their existence cannot be ruled out.

\vspace{0.3cm}

\subsection{Fast Rotation, Binarity, Flares}
Most of the stars with excess Nuv luminosity in Table~1 are associated with X-ray sources. This is a necessary but not sufficient
sign of stellar youth as active close binaries also possess elevated coronal X-ray emission \citep{mic}. The nearest (within 50 pc)
and the brightest X-ray emitters are phenomenologically separated into a few categories \citep{mak03} dominated by (1) RS CVn-type binaries
(with evolved components); (2) BY Dra-type active binaries (with MS components); (3) young stars; (4) contact binaries of WU UMa
type; (5) rapidly rotating single evolved stars. Short-period binaries feature strongly in this census with RS CVn pairs being the
most luminous X-ray emitters of all field non-degenerate stars. The fast rotation of components required to maintain high levels of
chromospheric and coronal activity is fueled by the angular momentum transfer via tidal interactions \citep{hut}. The same mechanism
relatively quickly circularizes tight orbits, but the presence of more distant, misaligned tertiary companions can be a source of
excitation for the eccentricity of the inner pair via the Lidov-Kozai cycle \citep{egg}. This probably explains the high rate of Rosat-detected
sources associated with resolved doubles \citep{mak02} -- these may be the visual components of interacting hierarchical triple systems.
A quarter of the sample have been detected as active binaries of BY Dra-type, often flaring and rotationally variable with structured photospheres.
Some objects of note include:
\begin{itemize}
\item
{\bf TYC 2331-1138-1 = CK Tri} is a variable mistakenly classified as RS CVn-type, but it is definitely a nearby pair of dwarfs of the
BY Dra-type. 
\item
{\bf HIP 3362 = FF And} is a BY Dra-type variable consisting of two twin M1V companions, also an astrometric binary with an orbital solution by
\citet{jan} with an orbital period of 2.170 d. \citet{chu} posited that the properties of the light curve are best explained by a large, 
cool spot on the surface. 
\item
{\bf HIP 51700} is one of the two F stars in the sample (F8), and possibly a short-period variable \citep{koe}. 
\item
{\bf HIP 45731 = GJ 3547} is a flare M1.0V star, which is a SB2 according to 
\citet{shk10} with an orbital period less than 20 d.
\item 
{\bf HIP 111802 = GJ 867A = FK Aqr}, a well-studied quadruple system of chromospherically active flare dwarfs.
The primary which is listed in Table~1 is a pair of twin dM1e stars \citep{her} with a period of 4.08 d.
\item
{\bf TYC 1355-214-1 = V429 Gem} is a K5Ve variable of BY Dra type, possibly including a brown dwarf companion \citep{he}.
\item 
{\bf HIP 14568 = GJ 3203 = AE For}, an eclipsing binary consisting of two K7Ve dwarfs with possibly a brown dwarf tertiary \citep{zas}.
\item
{\bf TYC 5858-1893-1} is one of the less studied stars of M2Ve type, detected as SB2 with a rotational and orbital period
of 2.9 d \citep{shk10}.
\end{itemize}

\vspace{0.3cm}

\section{A wider selection of UV-luminous stars}
A broader search for genuine hot stars in TGAS can be made if we drop the small distance criterion and consider the entire
population with statistically precise parallaxes and matching GALEX sources. Table~\ref{2.tab} lists 40 stars found with
the following criteria: $\varpi/\sigma_\varpi >5$, and Nuv$-G<2$ mag. The format of this table is the same as Table~1. 
Besides the previously found DN Dra, this selection includes one additional well-known WD of DA0.8 type, {\bf HIP~12031=FS~Cet}.
With $G=12.177$ mag, Nuv=12.371~mag, $M$nuv$=7.95$ mag, this star marks the top of the WD cooling sequence in the HR diagram.
Between FS Cet and van Maanen 2, the range of absolute Nuv magnitudes of white dwarfs is $\approx [8,20]$ mag, and this should make
the Nuv HR diagram a suitable proxy for the spectroscopic determination of type. 

\begin{table*}[!t]\centering
\begin{small}
  \setlength{\tabnotewidth}{\columnwidth}\tablecols{11}
  \setlength{\tabcolsep}{1\tabcolsep}
  \caption{Gaia stars with significant parallaxes and N\lowercase{uv}$-G$ colors less than 2 mag} \label{2.tab}
 \begin{tabular}{rrccccccccc}
    \toprule
\multicolumn{1}{c}{(1)} & \multicolumn{1}{c}{(2)} & (3) & (4) & (5) & (6) & (7) & (8) & (9) & (10) & (11) \\
\multicolumn{1}{c}{RA J2015} & \multicolumn{1}{c}{Dec J2015} & HIP & Tycho-2 & $\varpi$ & $\sigma_\varpi$ & $G$ & Fuv & Nuv & Fuv sig & Nuv sig \\
\multicolumn{1}{c}{$\deg$} & \multicolumn{1}{c}{$\deg$} & & & mas & mas& mag & mag & mag& mag & mag\\ \\
    \midrule
 57.3453 & 27.2266 & $$ & $1808-902-1$ & 1.94 & 0.27 & 11.49 & 12.453 & 13.337 & 0.005 & 0.004 \\
 60.133 & 27.4278 & $$ & $1821-1013-1$ & 2.43 & 0.3 & 11.425 & 12.827 & 13.321 & 0.005 & 0.004 \\
 26.1982 & 32.5499 & $$ & $2298-1538-1$ & 2.79 & 0.43 & 11.871 & 12.505 & 13.146 & 0.005 & 0.004 \\
 16.1481 & 41.2993 & $$ & $2807-1623-1$ & 5.41 & 0.55 & 13.128 & 13.489 & 13.859 & 0.009 & 0.007 \\
 143.672 & 30.561 & 46993 & $$ & 5.42 & 0.44 & 12.09 & 18.923 & 13.482 & 0.109 & 0.006 \\
 143.717 & 31.0274 & $$ & $2494-805-1$ & 4.67 & 0.56 & 12.849 & 15.752 & 13.734 & 0.026 & 0.006 \\
 154.487 & 55.2755 & $$ & $3818-1084-1$ & 1.68 & 0.26 & 11.656 & 11.903 & 12.634 & 0.004 & 0.004 \\
 111.785 & 26.9674 & $$ & $1918-1313-1$ & 1.67 & 0.3 & 11.888 & 13.728 & 13.711 & 0.008 & 0.004 \\
 102.816 & 56.6469 & $$ & $3774-18-1$ & 1.33 & 0.26 & 11.923 & 12.896 & 13.354 & 0.005 & 0.004 \\
 108.52 & 70.0716 & $$ & $4364-1209-1$ & 4.93 & 0.29 & 12.061 & 12.178 & 12.991 & 0.005 & 0.004 \\
 107.051 & 78.0469 & $$ & $4530-502-1$ & 1.35 & 0.23 & 12.052 & 12.719 & 13.075 & 0.004 & 0.004 \\
 246.563 & 23.0584 & $$ & $2043-1081-1$ & 2.3 & 0.45 & 10.847 & 12.1 & 12.558 & 0.005 & 0.003 \\
 252.108 & 59.0551 & 82257 & $$ & 91.04 & 0.5 & 12.288 & 13.606 & 13.443 & 0.007 & 0.004 \\
 216.785 & 72.9638 & $$ & $4416-1269-1$ & 2.04 & 0.33 & 11.126 & 11.737 & 12.195 & 0.003 & 0.002 \\
 350.178 & 38.1755 & $$ & $3230-1262-1$ & 3.86 & 0.31 & 12.875 & 19.167 & 14.12 & 0.077 & 0.005 \\
 300.943 & 71.6068 & $$ & $4454-1229-1$ & 3.23 & 0.3 & 10.434 & $$ & 12.039 & $$ & 0.002 \\
 13.0627 & -10.6629 & $$ & $5270-1692-1$ & 5.52 & 0.94 & 11.154 & $$ & 12.406 & $$ & 0.002 \\
 38.782 & 3.73248 & 12031 & $$ & 13.06 & 0.76 & 12.177 & $$ & 12.371 & $$ & 0.003 \\
 350.122 & 28.494 & $$ & $2249-1134-1$ & 2.22 & 0.31 & 11.83 & 12.966 & 13.39 & 0.005 & 0.003 \\
 0.551579 & 32.9799 & $$ & $2263-1340-1$ & 2.51 & 0.39 & 11.043 & 11.951 & 12.597 & 0.004 & 0.003 \\
 349.259 & 29.9058 & $$ & $2248-1765-1$ & 1.91 & 0.36 & 11.962 & 13.966 & 13.869 & 0.006 & 0.004 \\
 80.3045 & -24.7822 & $$ & $6479-610-1$ & 1.73 & 0.28 & 11.283 & 12.949 & 12.67 & 0.005 & 0.003 \\
 65.4181 & -6.01938 & $$ & $4733-1261-1$ & 3.27 & 0.34 & 11.385 & $$ & 13.334 & $$ & 0.004 \\
 235.694 & -7.72293 & $$ & $5597-9-1$ & 1.61 & 0.29 & 11.74 & 13.454 & 13.601 & 0.006 & 0.004 \\
 246.831 & 12.5777 & $$ & $967-861-1$ & 1.63 & 0.31 & 11.351 & 12.657 & 12.688 & 0.004 & 0.003 \\
 264.588 & 29.1466 & 86329 & $$ & 3.35 & 0.26 & 10.284 & 11.694 & 11.994 & 0.003 & 0.003 \\
 67.44 & -50.5233 & $$ & $8075-508-1$ & 2.68 & 0.25 & 11.758 & 18.242 & 13.599 & 0.094 & 0.006 \\
 75.8936 & -28.4547 & $$ & $6485-79-1$ & 1.66 & 0.28 & 12.249 & $$ & 12.812 & $$ & 0.004 \\
 27.1839 & -26.6038 & 8435 & $$ & 2.96 & 0.33 & 12.22 & 11.985 & 13.112 & 0.004 & 0.004 \\
 159.874 & -31.182 & $$ & $7186-829-1$ & 1.91 & 0.3 & 12.15 & $$ & 13.751 & $$ & 0.007 \\
 98.9003 & -62.6401 & 31481 & $$ & 2.45 & 0.44 & 12.348 & 12.906 & 13.351 & 0.004 & 0.003 \\
 95.8846 & -37.8134 & $$ & $7613-283-1$ & 2.2 & 0.28 & 11.203 & 12.529 & 13.006 & 0.004 & 0.003 \\
 294.23 & -59.285 & $$ & $8786-1818-1$ & 2.36 & 0.38 & 11.265 & 13.055 & 13.009 & 0.007 & 0.005 \\
 314.203 & -45.4108 & $$ & $8408-609-1$ & 1.63 & 0.31 & 12.314 & 12.532 & 12.925 & 0.006 & 0.002 \\
 349.927 & -55.6115 & $$ & $8834-986-1$ & 1.47 & 0.27 & 11.864 & 14.533 & 13.33 & 0.012 & 0.004 \\
 324.116 & -45.6489 & $$ & $8424-668-1$ & 8.2 & 0.3 & 13.304 & 22.525 & 15.158 & 0.405 & 0.009 \\
 325.129 & -31.4509 & $$ & $7487-82-2$ & 2.98 & 0.36 & 12.118 & 17.326 & 13.737 & 0.032 & 0.003 \\
 309.558 & -39.9754 & $$ & $7954-1134-1$ & 2.73 & 0.47 & 10.397 & 12.214 & 12.214 & 0.003 & 0.002 \\
 288.606 & -42.8892 & $$ & $7926-1427-1$ & 1.5 & 0.27 & 11.278 & 13.159 & 12.742 & 0.008 & 0.004 \\
 323.833 & -30.517 & $$ & $7474-402-1$ & 2.17 & 0.39 & 11.46 & 16.421 & 13.456 & 0.024 & 0.003 \\
    \bottomrule
  \end{tabular}
\end{small}
\vspace{0.4cm}
\end{table*}

Most of the objects in Table~\ref{2.tab} are relatively poorly studied stars that have remained under the radar of observers. It is only
now with the combination of precise GALEX photometry and Gaia parallaxes that we begin to see them as very unusual objects. Several
stars, on the contrary, have been studied in more detail, including:

\begin{itemize}
\item
{\bf TYC 2298-1538-1 = BG Tri} is a nova variable \citep{khr, kaz}.
\item
{\bf TYC 2807-1623-1 = RX And} is a famous dwarf nova \citep[e.g.,][]{kai}.
\item
{\bf TYC 5270-1692-1} has been previously recognized as a UV source owing to the observations with the TD1 satellite.
It is a binary with a solar-type and a hot subdwarf components (sdO$+$G)  \citep{ber}.
\item
{\bf HIP 86329} is another spectroscopically resolved solar-type -- hot subdwarf binary (sdOB$+$F/G)  \citep{ber}.
\item
{\bf TYC 6485-79-1} is a binary comprising a  solar-type and a hot subdwarf components (sdOB$+$F)  \citep{odo}.
\item
{\bf HIP 8435 = GJ 2026} is a solar-type -- hot subdwarf binary (sdO7$+$F/G)  \citep{gre}.
\item
{\bf HIP 31481 = RR Pic} is a nova variable \citep{sam}, and one of the earliest UV detections \citep{gal}.
\end{itemize}

The appearance of known novae and spectroscopic binaries with hot subdwarfs implies that more hidden WD and sdOB can be discovered
among the relatively nearby objects listed in Table~\ref{2.tab}. Follow-up spectroscopic and photometric observations are perhaps the best
way to find the nature of their excessive UV luminosity.

\vspace{0.3cm}

\section{Conclusions}
This study of nearby astrometric standards from the Gaia DR1 shows that stellar youth is only one of the
reasons for field stars to have excess Nuv luminosities, and perhaps, not the main one. Dynamical and possibly magnetic
interaction of low-mass dwarfs in close binaries is capable of supporting fast rotation rates and enhanced levels of
X-ray and UV radiation for durations comparable to the main-sequence lifetimes. This is confirmed, for example,
by in-depth investigations of stellar rotation rates in nearby open clusters. \citet{dou} find that almost all single
members of the Hyades (age $\approx 650$ Myr) with masses above $0.3\,M_{\sun}$ are slow rotators, while most of the
spectroscopic binaries in this mass range are fast rotators. Many of the nearby stars listed in Table~1 with Nuv$-G$ colors
bluer than the main sequence are expected to be old binary systems of BY Dra type. In a wider sample of stars with extreme
UV colors listed in Table~\ref{2.tab}, the presence of binaries with white dwarf and sdOB hot subdwarf components is conspicious,
but many others remain hidden.

It is reasonable to expect that metal-poor Population II stars should also show mild Nuv excess compared with disk dwarfs.
The weak absorption lines of metals in the near--UV region provide additional flux at short wavelengths. Since the fraction of Population II stars in the
solar neighborhood is low, we expect few, if any, such objects to be present in our analysis. Using tangential velocities
(computed from Gaia proper motions and parallaxes) as a proxy for population type, a search for high-velocity
stars within 40 pc of the Sun resulted in 70 objects (out of 1403) with $v_{\rm tan}>70$ km s$^{-1}$. These fast moving stars
comply with the main sequence quite well (not shown in this paper for brevity) with the exception of a few objects deviating
to the giant domain and possibly three stars with mild Nuv excess, all with $M{\rm nuv}$ around 15 mag. Only one of the three
objects satisfies the strict selection criteria adopted here, namely, the previously discussed spectroscopic binary HIP 45731,
but there is no evidence of metal deficiency in the literature.

It is also found that most of the field stars in the immediate solar neighborhood (distance less than 40\,pc) follow a well-defined and narrow
main sequence on the ``absolute Nuv magnitude versus Nuv$-G$ color" HR diagram constructed with Gaia parallaxes and GALEX and Gaia
photometry. This confirms the high quality of GALEX and Gaia photometric data and makes such a diagram a valuable method to detect
more stars with unusual UV radiation properties.

\acknowledgments
The author is grateful to J. Subasavage and J. Munn for useful discussions of the topic.
This work has made use of data from the European Space Agency (ESA)
mission {\it Gaia} (\url{http://www.cosmos.esa.int/gaia}), processed by
the {\it Gaia} Data Processing and Analysis Consortium (DPAC,
\url{http://www.cosmos.esa.int/web/gaia/dpac/consortium}). Funding
for the DPAC has been provided by national institutions, in particular
the institutions participating in the {\it Gaia} Multilateral Agreement. This research has made use of the VizieR catalogue access tool, CDS,
Strasbourg, France. The original description of the VizieR service was published in A\&AS 143, 23.

\end{document}